\documentclass[lettersize,journal]{IEEEtran}
\usepackage{amsmath,amsfonts}
\usepackage{algorithmic}
\usepackage{algorithm}
\usepackage{array}
\usepackage[caption=false,font=normalsize,labelfont=sf,textfont=sf]{subfig}
\usepackage{textcomp}
\usepackage{stfloats}
\usepackage{url}
\usepackage{verbatim}
\usepackage{graphicx}
\usepackage{cite}
\usepackage{float}
\usepackage{hyperref}
\begin{document}

\title{ Sound Analysis for Speed Estimation of Induction Motors Under Non-Stationary Conditions}

\author{Tomas A. Garcia-Calva,~\IEEEmembership{Member,~IEEE},
Daniel Morinigo-Sotelo, and Konstantinos N. Gyftakis,~\IEEEmembership{Senior Member,~IEEE}
\thanks{T. A. Garcia-Calva and Daniel Morinigo-Sotelo are with Research Groups ADIRE, ITAP, HSPDigital, and the Electrical Engineering Department of the University of Valladolid, 47001 Valladolid, Spain (e-mail: tom.garciacalva@ieee.org, daniel.morinigo@uva.es). (Corresponding author: T. A. Garcia-Calva)}
\thanks{K. N. Gyftakis is with the School of Electrical and Computer Engineering, Technical University of Crete, 73100 Chania, Greece (e-mail: k.n.gyftakis@ieee.org)}
\thanks{
\textcopyright~2026 IEEE. Personal use of this material is permitted.
Permission from IEEE must be obtained for all other uses, in any current
or future media, including reprinting/republishing this material for
advertising or promotional purposes, creating new collective works,
for resale or redistribution to servers or lists, or reuse of any
copyrighted component of this work in other works.
\vspace{0.5em}

This is the accepted version of the article published in
\textit{IEEE Transactions on Industry Applications}.
The early access version of record is available at:
\href{https://doi.org/10.1109/TIA.2026.3728739}
{https://doi.org/10.1109/TIA.2026.3728739}
}
}

\maketitle

\begin{abstract}

This paper presents a novel methodology for the estimation of the rotational speed of induction motors operating under non-stationary conditions, using acoustic signals acquired by a low-cost microphone. The proposed approach integrates multi-rate digital signal processing with advanced time–frequency analysis to extract speed-dependent harmonic components from motor acoustic emissions, while effectively suppressing electrical interference and electromagnetic noise.
Numerical simulations and experimental validations are conducted under both steady-state and transient operating conditions, considering a wide variety of speed profiles, including slow ramps and abrupt speed variations. The experimental results demonstrate that the proposed method is capable of accurately tracking rapid and dynamic speed changes caused by load disturbances and mechanical irregularities. 
The estimated instantaneous speed exhibits a high degree of agreement with reference measurements obtained from a conventional speed sensor.
The results confirm that the proposed acoustic-based technique provides a reliable, non-invasive, and cost-effective solution suitable for speed monitoring of induction motors.

\end{abstract}


\section{Introduction}
\IEEEPARstart{I}{nduction} motors (IM) play a key role in a wide range of industrial, transportation, and tertiary-sector applications, serving as the primary driving units in numerous processes due to their high efficiency and proven reliability \cite{pepe_review}. Their robust construction and capability to operate under diverse load conditions have made them the most widely deployed electrical machines worldwide. Consequently, ensuring efficient control, optimal operation, and accurate condition monitoring of induction motors is essential for maintaining productivity, operational safety, and system reliability across industrial environments. In this context, accurate speed estimation constitutes a fundamental requirement in induction motor control systems, as it directly enables effective torque and flux regulation \cite{antonySensorlessFieldOriented2022}. Moreover, precise instantaneous speed information is essential for reliable condition monitoring and fault diagnosis. This is particularly important because many fault-related signatures manifest as speed-dependent harmonic components in both mechanical and electrical variables, such as vibration signals and stator currents. Consequently, high-accuracy speed estimation is critical for ensuring stable motor operation and for enhancing the effectiveness of health monitoring and fault detection strategies.

The speed of induction motors can be accurately measured using modern sensors such as resolvers, tachometers, and encoders \cite{tii_giorgos, muzioDetectionBrokenBars2024}. Although these devices provide highly precise measurements, their implementation is often associated with significant cost and practical limitations. Their installation is intrusive and typically requires mechanical modifications to the motor or shaft. Furthermore, they are subject to mechanical wear and demand periodic maintenance, which increases operational costs and may compromise system reliability. In many industrial environments, harsh operating conditions, such as elevated temperatures, dust, humidity, and mechanical vibrations, can further degrade the accuracy, reliability, and lifespan of conventional speed sensors \cite{Tenergies25, germenSoundBasedInduction2014}.

Sensorless speed estimation techniques have therefore emerged as a compelling alternative to mechanical sensing devices \cite{abb_framework, elbarbaryReviewSpeedEstimation2024}. These approaches rely on the analysis of measurable motor signals, including stator voltages and currents \cite{mie2011briz}, mechanical vibrations \cite{peetersReviewComparisonTacholess2019, urbanekTwostepProcedureEstimation2013}, and electromagnetic torque. By exploiting these signals through mathematical modeling \cite{garcia-calvaRotorSpeedEstimation2024} and advanced signal processing, the rotor speed can be inferred without the need for direct physical sensors.

Each type of signal used for speed estimation presents specific advantages and limitations. Voltages and stator currents are widely employed for condition monitoring and can also be utilized for speed estimation. Their implementation is relatively straightforward, as stator current measurements are typically readily available. However, the accuracy of current-based speed estimation can be affected by load variations and the presence of interference harmonics \cite{elbarbaryReviewSpeedEstimation2024}. Mechanical vibration signals contain direct information related to motor dynamics and can be exploited for rotational speed estimation \cite{urbanekTwostepProcedureEstimation2013,sarrio}. However, vibration-based measurements generally rely on contact sensors, such as accelerometers, whose installation may be challenging in certain motor configurations. Moreover, the measured vibration amplitudes are strongly influenced by mounting conditions, surface characteristics, and the mechanical structural properties of both the motor and the driven load.

To overcome the limitations associated with current and vibration-based speed estimation methods, non-invasive monitoring strategies based on acoustic signatures have recently attracted increasing attention \cite{bankovicSoundbasedSpeedEstimation2024,kliemankInstantaneousAngularSpeed2023, alshormanSoundsAcousticEmissionbased2021,whirpool,avyner}. Acoustic measurements do not require physical contact with the machine, thereby avoiding intrusive sensor installation. Moreover, the sound emitted by an induction motor is closely related to its electromechanical behavior, enabling accurate speed estimation. In addition, sound sensors (microphones) are commercially widespread and less expensive than specialized speed, current, and vibration sensors, which further supports the practicality and scalability of acoustic-based monitoring approaches.
These characteristics make acoustic-based techniques particularly attractive for low-cost and flexible monitoring of induction motors in industrial applications \cite{rajapakshaAcousticAnalysisBased2021}.

Despite these advantages, extracting reliable and accurate speed information from acoustic signals is a challenging task. The presence of multiple harmonic components, strong noise, and non-stationary operating conditions often leads to spectral smearing and limited frequency resolution when conventional FFT-based methods are applied. Furthermore, many existing studies focus primarily on steady-state operation or averaged spectral representations, which are insufficient for instantaneous speed tracking under dynamic conditions. To address these issues, time–frequency analysis techniques, such as the Short-Time Fourier Transform (STFT) and wavelet-based methods, have been employed to capture the time-varying characteristics of acoustic signals \cite{urbanekTwostepProcedureEstimation2013}. Recent works have demonstrated the potential of advanced techniques for instantaneous speed estimation in IM operating under variable conditions. For instance, FFT-based acoustic emission analysis has been explored for angular speed estimation in gas foil bearings \cite{kliemankInstantaneousAngularSpeed2023}, while subspace-based methods, such as Singular Value Decomposition (SVD), have been successfully applied to enhance the extraction of speed-dependent harmonics under variable-speed operation \cite{tom2}. These approaches highlight the importance of high-resolution processing tools capable of tracking frequency variations over time \cite{combetNewMethodEstimation2009, toutiImprovedElectromechanicalSpectral2018}.

In this work, a methodology for the estimation of IM speed using only acoustic measurements is proposed. This work extends the approach previously introduced in \cite{Tsdemped25}, where preliminary results on acoustic-based speed estimation were presented. The method exploits the harmonic structure associated with the blade passing frequencies (BPF) and incorporates a multi-rate signal processing strategy to isolate the frequency bands of interest with improved computational efficiency. 
A Short-Time Multiple Signal Classification (ST-MUSIC) algorithm is then employed to achieve high-resolution frequency estimation, enabling accurate speed tracking under load variations and non-stationary operating conditions. The proposed approach is studied through numerical simulations and validated by experimental tests conducted on a dedicated test bench, where the estimated speed is quantitatively compared with measurements obtained from a reference tachometric sensor. 
The results demonstrate that the proposed acoustic-based method provides accurate and robust speed estimation, highlighting its potential as a non-contact, low-cost, and easily deployable solution for industrial monitoring.

\section{Mathematical Foundations}

\subsection{Induction Motor Speed}
The three-phase asynchronous motor is widely used in many sectors. Its operating principle is well known. When the stator winding is fed with a balanced power supply, it creates a constant rotating magnetic field in the machine's air gap. This magnetic field rotates at the synchronous speed $(n_s)$, which depends on the voltage supply frequency $(f_s)$ and the number of pole pairs of the machine $(p)$:

\begin{equation}
n_s = \frac{f_s}{p}
\label{eq_n1}
\end{equation}

According to Faraday's law, currents circulate through the rotor and, together with the magnetic field in the air gap, produce an electromagnetic torque that causes the rotor to rotate in the same direction as the rotating magnetic field. However, the rotor speed never reaches the synchronous speed, and the difference between synchronous and rotor speeds is defined as the slip $(s)$:
\begin{equation}
s = \frac{n_s - n_r}{n_s}
\label{eq_s}
\end{equation}
\noindent where $n_r$ is the rotor speed. The mechanical rotor speed in revolutions per second (\text{r.p.s.}) can be expressed as
\begin{equation}
n_r = (1-s)\frac{f_s}{p}.
\label{eq_nr}
\end{equation}

When the machine operates under no-load conditions, the rotor speed approaches the synchronous speed. In steady-state operation, the electromagnetic torque generated by the motor balances the load torque. As the load torque increases, the rotor speed deviates further from the synchronous speed, resulting in a higher slip. If a sudden variation in the load torque occurs, the motor torque responds almost instantaneously, while the rotor speed undergoes a transient response until a new steady-state condition is reached. The duration of this transient depends primarily on the equivalent moment of inertia of the drive system.
The mechanical behavior of the IM following a disturbance in the load torque can be described by the rotational form of Newton’s second law. The dynamic equilibrium of the drive system is governed by the balance between the electromagnetic torque developed by the motor, the load torque, and the mechanical losses, while accounting for the inertia of the rotating masses. This relationship can be expressed in expanded form as
\begin{equation}
J \frac{d\omega(t)}{dt} + B\,\omega(t) = T_e(t) - T_L(t),
\label{eq_mech_dyn}
\end{equation}
where $J$ denotes the equivalent moment of inertia of the motor–load system, $\omega(t)$ is the angular speed of the rotor, $B$ represents the viscous friction coefficient, $T_e(t)$ is the electromagnetic torque, and $T_L(t)$ is the load torque.

\subsection{Short-Time Analysis Using Subspace Decomposition}  \label{sta}

This study presents an approach for estimating the rotational
speed of an IM by evaluating its emitted sound. Load torque disturbances, governed by \eqref{eq_mech_dyn}, manifest acoustically as time-varying harmonic components in the motor's radiated sound, requiring a processing approach capable of jointly resolving spectral content and its temporal evolution. To this end, the subspace-based Short-Time Multiple Signal Classification (ST-MUSIC) technique \cite{tec_2019} is employed to track these non-stationary harmonic characteristics through high-resolution time-frequency (t-f) analysis. The power spectral density $(P_{xx})$ of the $N$-observation samples of the sound signal is defined as the discrete-time Fourier transform of its autocorrelation sequence::

\begin{equation}
P_{xx}(f) = T \sum_{k=-\infty}^{+\infty}{r_{nn}[k]e^{-j2\pi fkT}}
\label{eq_psd}
\end{equation}
\noindent where the variable $r_{nn}[k]$ is the autocorrelation function of $s_r[k]$, defined as $r_{nn}[k] = E[s_r[l] s_r[k+l]]$. The autocorrelation matrix of the sound signal sequence $s_r[k]$ can be written as a sum of other autocorrelation matrices and noise $\eta$: $R_{nn} = R_{xx} + R_{\eta\eta}$. The spectral estimation of the sound signal is:

\begin{equation}
\hat{P}_{xx}(f) = \frac{1}{s^H(f)V(f)V^H(f)s(f)}
\label{eq_p}
\end{equation}
\noindent where $V$ is the matrix of eigenvectors of the noise subspace and $s$ is a vector of complex sinusoids \cite{tec_2019}. The resulting spectrum displays sharp peaks at frequencies of the sound signal oscillations. Hence, this power spectral density estimation is used for a high-resolution time-frequency analysis.

\section{Blade Passing Frequencies}

In rotating machinery such as compressors, turbines, and motors, Blade Passing Frequencies (BPF) arise as a direct consequence of periodic aerodynamic interactions between the rotating parts and the air flow. When a fan with $N_b$ blades rotates at an angular speed $\omega_r$ (rad/s) or at a rotational frequency $f_r = \omega_r / 2\pi$, each blade periodically disturbs the air flow, producing a fluctuating pressure field. The periodic disturbance yields a base frequency given by $f_{\text{B}} = N_b f_r$. 
The total acoustic pressure signal $p(t)$ can therefore be modeled as a periodic repetition of the single blade pass

\begin{equation}
p(t) = \sum_{k=-\infty}^{\infty} s\big(t - kT\big)
     = s(t) * \sum_{k=-\infty}^{\infty} g(\tau - T_{\text{B}}),
\end{equation}

where $T_{\text{B}} = 1/f_{\text{B}}$, and $g(t)$ is the pressure signature from a single blade passage of width $\tau$. In the frequency domain, the periodic structure of \(p(t)\) leads to a discrete spectral distribution dominated by the base frequency $f_{\text{B}}$ and its integer harmonics. Specifically, the spectral analysis of the periodic blade-pass waveform produces spectral lines at \( f_{\text{BPF} } = m f_{\text{B}}\), with $m \in \mathbb{Z}^+$, 
whose amplitudes follow \cite{vetterli}
\begin{equation}
    |G(m f_{\mathrm{B}})| = A \tau \left|\mathrm{sinc}(\pi m f_{\mathrm{B}} \tau)\right|,
    \label{eq_envelope}
\end{equation}
producing the characteristic BPF harmonics commonly observed in airborne acoustic spectra.

\medskip
When the rotating element is driven by an IM, the blade passing frequencies become directly linked to the rotor speed of the machine. According to~\eqref{eq_nr}, the resulting BPF harmonics can be expressed as
\begin{equation}
f_{\mathrm{BPF}} = \frac{m N_b}{p} (1 - s) f_s, \qquad m \in \mathbb{Z}^+.
\label{eq_n}
\end{equation}
These harmonics represent the complete BPF family. In practical IM systems, however, the acoustic field generated by the motor's fan is not uniformly distributed across all BPF components, but is governed by the aerodynamic symmetry imposed by the number of blades $N_b$ and their spatial arrangement. This symmetry leads to selective reinforcement or cancellation of specific harmonics due to constructive or destructive interference. In particular, the fundamental blade passing frequency $f_{\text{B}}$ and its prime harmonics typically exhibit higher amplitudes, as they arise from the superposition of pressure fluctuations generated by successive blade passages and are less affected by phase cancellation mechanisms. Let $\mathbb{P}$ denote the set of prime numbers and $p_k$ the $k$-th prime element. The blade passing harmonics associated with prime orders can be expressed as
\begin{equation}
f_{B,p_k} = p_k \, f_B, \quad p_k \in \mathbb{P},
\label{eq:prime_bpf}
\end{equation}
where $f_B$ is the fundamental blade passing frequency.

The predominance of prime-order BPF components stems from the aerodynamic symmetry and spatial periodicity of the blade arrangement. Composite orders $m = a \cdot b$ (with $a,b>1$) share common divisors with $N_b$, causing partial cancellation through correlated, phase-aligned pressure fluctuations across rotationally symmetric sub-arrays of blades. Prime orders, lacking such divisors, avoid this phase alignment and preserve reinforcement over successive revolutions, yielding higher radiated amplitudes. For prime blade counts (e.g., $N_b = 7$), where no sub-array symmetry exists, this reinforcement instead arises from a second-order nonlinear mechanism: the quadratic convective term of the flow generates combination tones at the sum and product frequencies of lower-order components~\cite{schetzen2006}, preferentially coupling composite orders while leaving prime orders, admitting no such factorization, governed solely by the sinc envelope of Eq.~\eqref{eq_envelope}. Prime-order BPF harmonics thus constitute the most reliable spectral markers for speed estimation, exhibiting higher signal-to-noise ratio (SNR) under both stationary and non-stationary conditions.

\section{Proposed Methodology}

\begin{figure}[htbp]
\centerline{\includegraphics[width=0.45\textwidth]{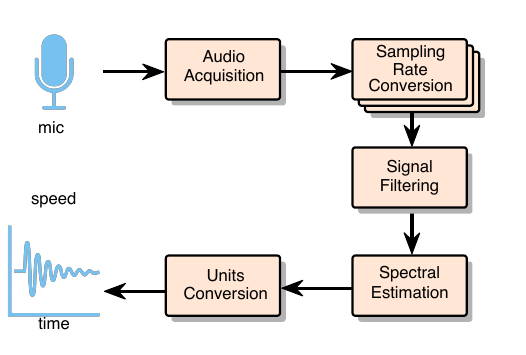}}
\caption{Block diagram of the proposed motor speed estimation method.}
\label{fig_pm}
\end{figure}

This section presents the proposed methodology for estimating the instantaneous speed of IM based on the analysis of acoustic signals emitted during operation. The method comprises five stages: signal acquisition, sampling rate conversion, signal filtering, short-time analysis, and frequency-to-speed conversion. The complete signal processing chain is illustrated in Fig.~\ref{fig_pm} and is described as follows:

\subsection{Audio Acquisition}
The acoustic signal is captured by a microphone, which converts sound pressure variations into a voltage signal. To prevent aliasing, the signal is passed through an analog low-pass filter with a cutoff frequency of 30~kHz prior to digitization. The signal is then sampled at
$f_n=80$~kHz, a rate selected to emulate the acquisition characteristics of commercial audio interfaces (typically 44.1--192~kHz), thereby enabling implementation of the proposed method with low-cost, off-the-shelf audio hardware while fully exploiting the microphone's frequency response through oversampling.

\subsection{Sampling Rate Conversion}

The discretized acoustic signal is preprocessed using a multi-rate architecture designed to isolate frequency components within the range $[0, 2)\,\text{kHz}$. The preprocessing chain consists of a digital low-pass (LP) filtering stage followed by a down-sampling stage with a factor of $M = 20$. Prior to down-sampling, the signal is passed through the LP filter with cutoff frequency $f_c = f_n/2M$, ensuring that no spectral components above the Nyquist frequency of the decimated signal are present, thereby preventing aliasing in accordance with the sampling rate conversion theory~\cite{proakis2007}. This effectively reduces the sampling rate from $f_n = 80\,\text{kHz}$ to $f_n' = 4\,\text{kHz}$, while preserving the relevant signal components within the target band. The multi-rate architecture ensures sharp transitions between the pass-band and stop-band, avoiding amplitude distortion within the pass-band region.

\subsection{Signal Filtering}

After sampling rate conversion, the signal has a new sampling period of $250~\mu$s, corresponding to a bandwidth (BW) of $[0, 2)\,\text{kHz}$. To further refine the signal, a high-pass (HP) filter is applied to remove low-frequency components unrelated to the motor’s rotational speed. This stage isolates the higher-order harmonic of interest, specifically the 11th BPF harmonic, which carries the speed information while avoiding the low-frequency region that is typically contaminated by dense electromagnetic noise. The HP filter is implemented as an 8th-order IIR Butterworth filter using second-order sections, with a steepness factor of 0.85 and a stopband attenuation of 60~dB. This configuration ensures effective suppression of undesired low-frequency components while preserving the amplitude characteristics of the harmonic of interest.

\subsection{Short-Time Analysis}

With the signal's spectral content confined to a narrow BW, the next step involves segmenting the audio signal for time-localized spectral analysis. The signal is divided into overlapping short-time segments using a rectangular window of 256 samples. Each segment is independently transformed into the frequency-domain using the high-resolution estimator described in ~(section~\ref{sta}). The adoption of a subspace-based estimator over classical methods is motivated by its superior frequency resolution. FFT-based estimators are constrained by the Rayleigh resolution limit ($\approx 1/NT$), which restricts their ability to resolve spectral components for a short observation length $N$. In contrast, subspace decomposition exploits the eigenstructure of the autocorrelation matrix $R_{nn}$, separating the signal and noise subspaces to achieve asymptotically unbiased, high-resolution estimates of the number and frequency of spectral components, even in the presence of noise \cite{stoica1989}. This estimator is applied over short-time segments of the signal, producing a time–frequency representation that captures how spectral components evolve.

\subsection{Frequency-to-Speed Conversion}

After the application of the frequency estimation algorithm for harmonic decomposition, the frequency axis of the resulting representation is transformed into rotational speed units (revolutions per second). This mapping relies on the identification of the frequency location of the 11th blade BPF harmonic at each short-time segment, which is directly associated with the instantaneous rotational speed of the motor. 
Once the frequency of this harmonic is estimated using (\ref{eq_nr}), (\ref{eq_n}), and (\ref{eq:prime_bpf}), the rotational speed is obtained by normalizing the instantaneous frequency with respect to both the harmonic order and the number of blades, i.e., $\hat{n}_r(t) = f_{B,11}(t)/(11\,N_b)$. 
This transformation provides a direct representation of the motor's time-varying speed profile.

The proposed signal chain enables speed monitoring of IM based on sound analysis and under dynamic conditions. The combination of multi-rate processing and signal filtering allows spectral segmentation and ensures that the signal content is concentrated in the relevant harmonic band associated with BPF harmonics. Finally, the application of the high-resolution algorithm improves the robustness of the spectral estimation, effectively distinguishing harmonic components from noise, even in low SNR environments. This robustness makes the approach particularly well-suited for scenarios involving weak signals and noisy measurements, where conventional spectral estimation techniques typically underperform.

\section{Numerical Results}

Numerical simulations were conducted considering a two-pole-pair squirrel-cage induction motor rated at 18.45 kVA, 400 V AC, 50 Hz. The total inertia, including the rotor and mechanical load, was set to $J = 1.0\ \mathrm{kg\ m^2}$, while viscous friction was set to $B = 0.0001\ \mathrm{Nm\ s}$. The motor was modeled using per-phase parameters and equivalent-circuit representations, with both electrical and mechanical dynamics incorporated to capture transient and steady-state behaviors accurately \cite{garcia-calvaRotorSpeedEstimation2024}. 

The simulation model employed in this work targets the spectral structure of the fan's acoustic emission, focusing on the BPF and its harmonic series rather than absolute sound-pressure levels or the underlying turbulent flow field. Since $f_{\text{BPF}}$ is a kinematically determined quantity, fixed by the blade count $N_b$ and shaft speed, it can be predicted from a rotating-force (loading) and thickness-source representation without resolving the full turbulent flow. This is supported by classical analyses of rotating-source acoustics~\cite{lowson1965}, which show that a rotating aerodynamic force generates discrete tones at integer multiples of the BPF, and by the Ffowcs Williams–Hawkings acoustic analogy~\cite{fwh1969}, which establishes that, for a compact, low-Mach-number fan, the tonal field is dominated by the thickness and loading (monopole/dipole) terms, steady in blade-fixed coordinates, while the quadrupole (turbulence) term—requiring full computational aeroacoustics—remains negligible for this purpose. This is further corroborated by reduced-order, frequency-domain harmonic-noise theory~\cite{hanson1980}, shown to yield engineering-accuracy predictions of the harmonic noise structure without a full aeroacoustic multiphysics simulation. Accordingly, the electromagnetic motor model, which fixes the instantaneous shaft speed, is coupled to a signal/systems synthesis of the BPF harmonic series—a physically grounded and sufficient approach for validating speed estimation from the spectral position of prime-order BPF harmonics.

\begin{figure}[htbp]         
\centerline{\includegraphics[width=0.46\textwidth]{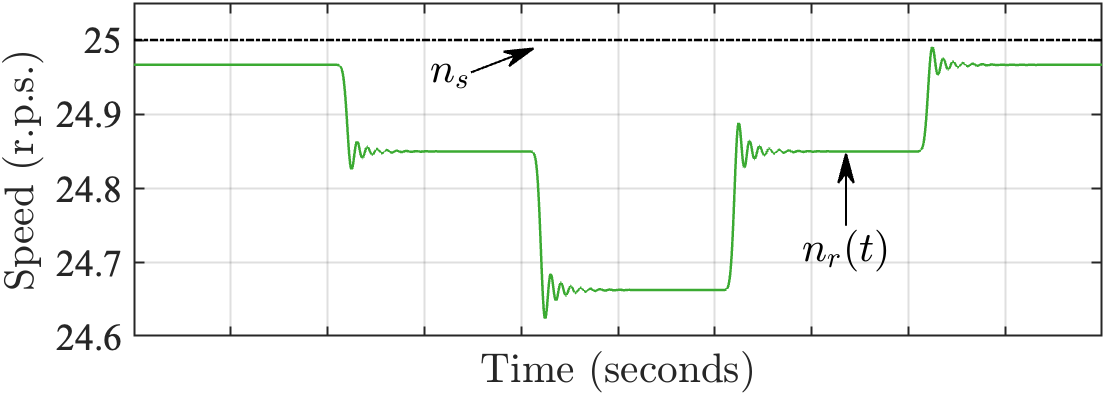}}
\caption{Numerical Results. Speed response to step load torque variations; x-axes: Time 1 second/div.}
\label{fig_speed}
\end{figure}

The simulations were carried out considering dynamic changes in the motor speed. As described in Section III, variations in the load torque modify the slip value, and consequently the motor speed. These variations directly affect the power spectral density content of the motor-generated sound, especially the $f_{\text{BPF}}$. The load torque was modeled as time-dependent step functions, imposing abrupt changes in the mechanical resistance applied to the motor shaft. The resulting BPF were simulated in the time-frequency domain.  Fig.~\ref{fig_speed} shows the dynamic behavior of the motor  speed under the load torque variations. During the first 5 seconds, two positive torque steps were applied: the first of amplitude $1$~Nm at $t=2$~s, and the second of $2.6$~Nm at $t=4$~s. 

These disturbances increased the developed load torque, causing a noticeable reduction in the rotor speed as the slip increased. From $t=5$~s to $t=10$~s, two negative torque steps of the same absolute amplitudes were applied at $t=6$~s and $t=8$~s, respectively. In these intervals, the decrease in the load torque allowed the rotor to accelerate, moving its operating point closer to the synchronous speed ($n_s$). Due to motor inertia, a transient response preceded the new steady-state speed after each load change.

\begin{figure}[htbp]
\centerline{\includegraphics[width=0.46\textwidth]{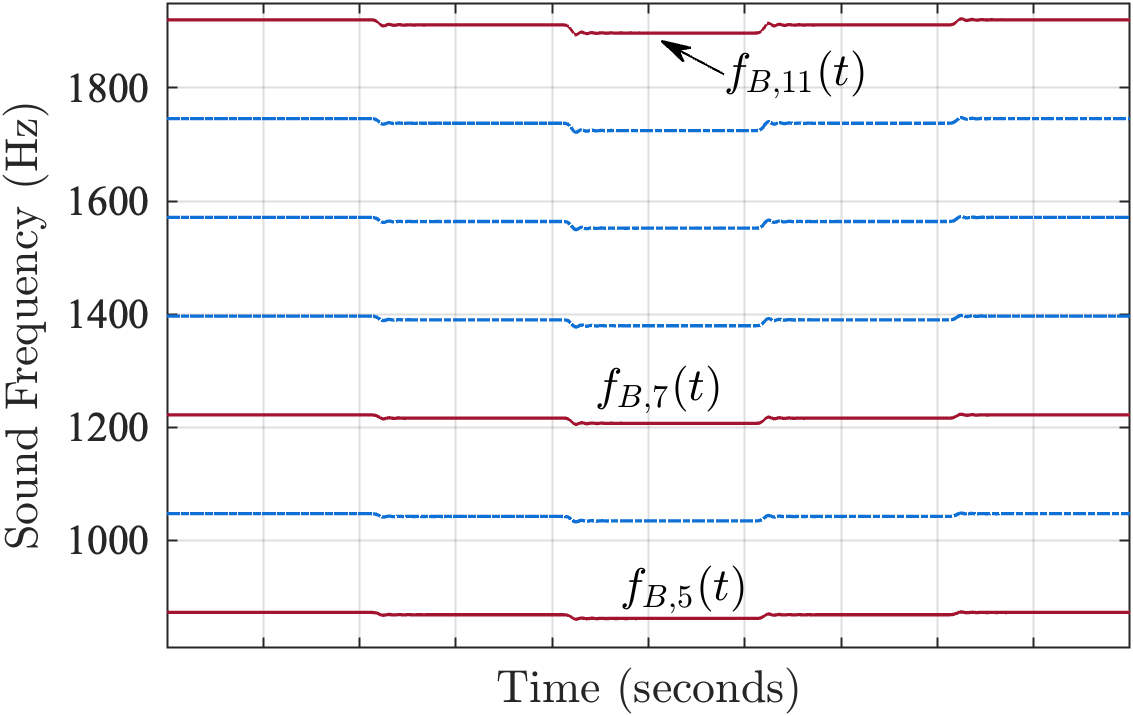}}
\caption{Time-frequency blade pass harmonics; x-axes: Time 1 second/div.}
\label{fig_bpf}
\end{figure}

Fig.~\ref{fig_bpf} illustrates the BPF, from the $f_{B,5}$ to the $f_{B,11}$ order, observed over 10 seconds of the simulation time within the frequency band from 900 to 1925~Hz. Although the acoustic emission of the motor-driven fan contains the complete BPF family, the dominant spectral components correspond to the prime-order harmonics, which preserve significantly higher amplitudes.  Variations in rotor speed induce frequency shifts in the BPF components, as evidenced by the time-frequency representation. Specifically, an increase in load torque results in a downward frequency shift of the BPF family due to a reduction in rotor speed, whereas a decrease in load torque produces an upward shift as the rotor accelerates and the slip is reduced. Furthermore, higher-order BPF components demonstrate increased sensitivity to speed variations, reflecting pronounced frequency modulation effects directly linked to rotor speed dynamics. These frequencies are fundamentally determined by the product of the rotational speed and the number of blades. 
The 11th harmonic illustrates the variations in the time-varying acoustic spectral features as changes in motor loading and rotational speed occur.  
When a resistive torque is applied to the motor (2 sec), it undergoes a deceleration that opposes the inertia of motion, leading to a corresponding decrease in the frequency of the emitted acoustic harmonic. Conversely, during acceleration (6 sec), the frequency of the emitted sound increases proportionally with the rotational speed.
For an axial fan with $N_b = 7$ blades operating at $25~\text{r.p.s.}$, the rotational frequency is $f_r = 25~\text{Hz}$, and the fundamental BPF component is given by $f_{B,1} = N_b f_r = 175~\text{Hz}$. Accordingly, the dominant component appears at $175~\text{Hz}$, with higher-order harmonics at integer multiples (e.g., $350$, $525$, $875$, $1225$, and $1925~\text{Hz}$).

The 11th BPF harmonic was selected as the speed feature because it falls within the effective frequency response of the low-cost acoustic sensor, while offering a favorable trade-off between speed sensitivity, which increases with harmonic order \eqref{eq:prime_bpf}, and harmonic amplitude, which decreases with harmonic order \eqref{eq_envelope}. This order provides both a detectable amplitude and adequate sensitivity to rotor speed variations.

\section{Test Bench}

\begin{figure}[htbp]
\centerline{\includegraphics[width=0.46\textwidth]{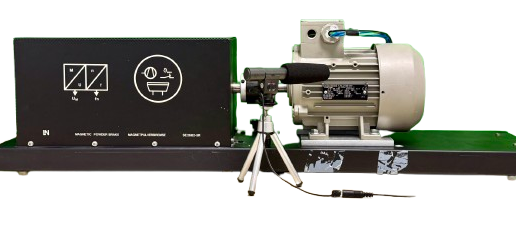}}
\caption{Laboratory test bench for acoustic-based speed estimation.}
\label{fig_laboratory}
\end{figure}

The laboratory setup detailed below is illustrated in Fig. \ref{fig_laboratory}. This laboratory configuration includes a three-phase induction motor, a controllable magnetic brake, an LP filter, a data acquisition board, and a laptop. The three-phase induction motor is a Siemens model 1LA7083-4AA10 with the stator winding in Y-configuration. This motor has the following specifications: rated power of 750 W, rated voltage of 400 V, and rated current of 1.86 A. The motor is coupled with a Lucas-Nülle SE 2662-5R brake, which offers a braking power of 1 kW and allows manual or voltage-controlled torque settings. This brake also provides a tachometer and a torque sensor. Motor sound was measured with a low-cost Shenggu SG-108 microphone model. The unidirectional microphone is employed to capture the acoustic signals generated by the target motor, while effectively attenuating noise from surrounding machinery. This directional characteristic ensures that the recorded signal predominantly contains information of interest. The sound signal passes through the LTC 1564 filter. Signal acquisition is facilitated by a NI system comprising the cDAQ-9174 chassis and the NI-9215 module. This system connects through a USB port to the laptop. Data acquisition and processing are performed directly through MATLAB software. The sampling frequency was 80 kHz. Multiple experiments were conducted with the motor powered by the mains at 50 Hz, operating under stable conditions and experiencing load variations that altered its speed.

\section{Experimental Results}

\begin{figure}[htbp]
\centerline{\includegraphics[width=0.46\textwidth]{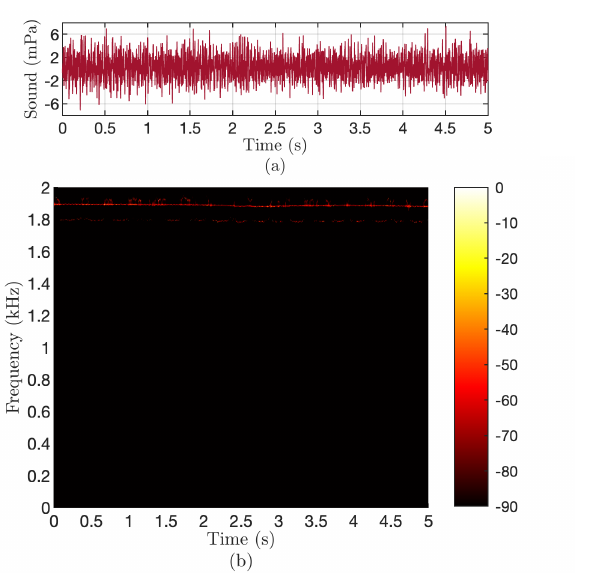}}
\caption{ Experimental results: (a) Time-domain waveform with original bandwidth, and (b) Time-frequency decomposition of the processed sound signal. Frequency bandwidth = $[0, f_n/2M)$.}
\label{fig_er1}
\end{figure}

The experiments involve the analysis of the acoustic signal captured by the microphone while the motor operates under different load and operating conditions. Fig.\ref{fig_er1}(a) presents the time-domain waveform of the sound. As depicted, the signal exhibits significant amplitude fluctuations, corresponding to pressure variations detected by the microphone sensor, which reflect the dynamic behavior of the sound. However, the waveform alone poses substantial challenges for directly extracting information related to the instantaneous speed due to the complex behavior of the full-band signal and the presence of environmental and measurement noise. 

In contrast, Fig.~\ref{fig_er1}(b) shows the spectrogram of the sound signal after processing through the first four stages of the proposed methodology. This representation illustrates the effectiveness of the approach in constraining the spectral content of the signal and improving its robustness to environmental noise. Specifically, the initial stage of sample rate conversion successfully eliminates spectral density beyond 2~kHz, thereby reducing computational burden and focusing the analysis on the most informative BW. Subsequently, the filtering stage removes the spectral components within the [0, 1.7)~kHz range. The resulting signal is further constrained through subspace decomposition, which benefits from the prior frequency-band conditioning to isolate and retain only the dominant high-amplitude components associated with the induction motor’s operational dynamics while effectively suppressing measurement noise. This sequential processing enables a more precise localization of the speed-related component embedded within the acoustic signal.

\begin{figure}[htbp]
\centerline{\includegraphics[width=0.46\textwidth]{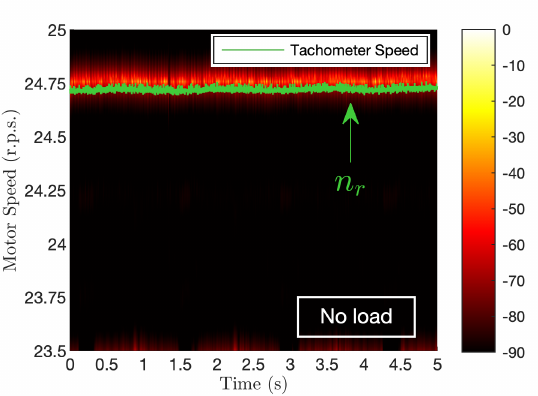}}
\caption{Experimental results: induction motor operating in no-load conditions.}
\label{fig_er2}
\end{figure}

\begin{figure}[htbp]
\centerline{\includegraphics[width=0.46\textwidth]{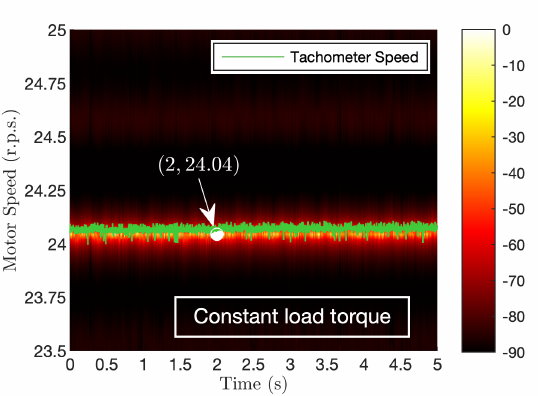}}
\caption{Experimental results: induction motor operating under 2 Nm controlled load torque.}
\label{fig_er3}
\end{figure}

Figs.~\ref{fig_er2} and~\ref{fig_er3} illustrate the performance of the full proposed methodology for motor speed estimation under steady-state operating conditions. The time-speed representation is in the speed range of [23.5, 25]~r.p.s, corresponding to full-load and no-load, respectively. Specifically, Fig.~\ref{fig_er2} presents the short-time decomposition of the acoustic signal over a 5-second interval, whereas the induction motor operates without any mechanical load coupled to the rotor shaft. 
 The analysis enables the estimation of an average speed of 24.75~r.p.s. Notably, the proposed method reveals a rotor speed close to the synchronous speed (25 r.p.s.), minor low-frequency fluctuations, and slight ripples in the speed behavior, which can be attributed to inherent imperfections in the motor's geometry. Although no external load variations are present, the operating condition is better described as pseudo steady-state due to these intrinsic periodic and aperiodic speed deviations. In contrast, Fig.~\ref{fig_er3} depicts the speed estimation outcome when the motor operates under full load, imposed by a uniformly controlled torque using the electromagnetic brake. The method successfully estimates a stable speed of approximately 24.044~r.p.s. at the 2.0-second mark, maintaining a consistent behavior throughout the 5 displayed seconds. Importantly, these steady-state experimental results confirm that the speed-dependent sound harmonic resides within an appropriate frequency range, effectively avoiding interference from overlapping sound harmonics. This characteristic facilitates reliable speed monitoring. Although the aforementioned results provide a useful reference for analysis, the behavior observed does not fully represent real industrial operating conditions. In practice, motors are rarely operated under no-load or full-load conditions, as such regimes reduce efficiency and may lead to damage. Furthermore, motor operation is not restricted to steady-state conditions; most industrial motors operate under dynamic regimes, spanning from minimum to maximum load, and are subject to transient variations in load levels.

\begin{figure}[htbp]
\centerline{\includegraphics[width=0.46\textwidth]{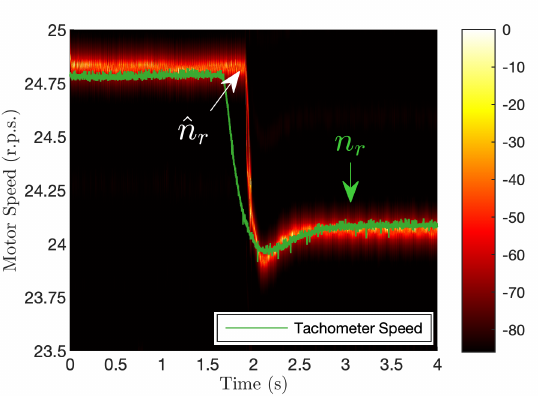}}
\caption{Experimental results: induction motor operating under a positive step load torque.}
\label{fig_pos_torqueStep}
\end{figure}

Fig.~\ref{fig_pos_torqueStep} illustrates the transient response of the estimated speed under dynamic conditions when a positive load torque is applied at $1.9$ seconds. Initially, under no-load operation, the estimated speed remains constant at 24.81~r.p.s. When the load torque is applied, a fast transient response is observed, characterized by an abrupt decrease in speed to 23.88~r.p.s. at 2.09~sec. Subsequently, the system converges to a new steady-state operating point at 24.07~r.p.s.

\begin{figure}[htbp]
\centerline{\includegraphics[width=0.46\textwidth]{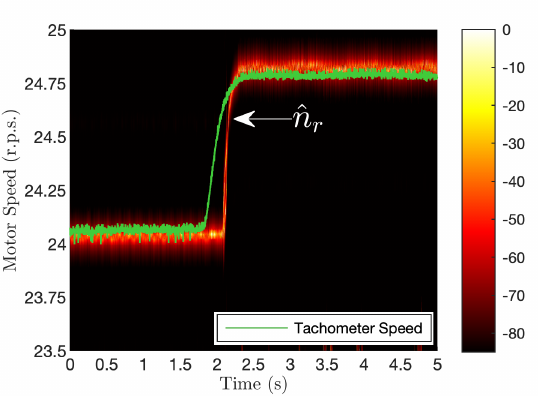}}    
\caption{Experimental results: induction motor operating under a negative step load torque.}
\label{fig_neg_torqueStep}
\end{figure}

The results depicted in Fig.~\ref{fig_neg_torqueStep} illustrate the performance of the proposed speed estimation method under a load removal condition. Initially, the induction motor operates under a constant load, resulting in a steady-state estimated speed of 24.04~r.p.s. At $t = 2.1$~s, the mechanical load is abruptly removed, leading to a transient acceleration phase clearly tracked by the method. The estimated speed rapidly increases and settles around a new steady-state value of 24.81~r.p.s. Notably, the proposed method accurately tracks the dynamic transition, exhibiting a fast response in both state regimes, indicating strong robustness against noise and the capability of the estimator to reliably follow non-stationary operating conditions. 

\begin{figure}[htbp]
\centerline{\includegraphics[width=0.46\textwidth]{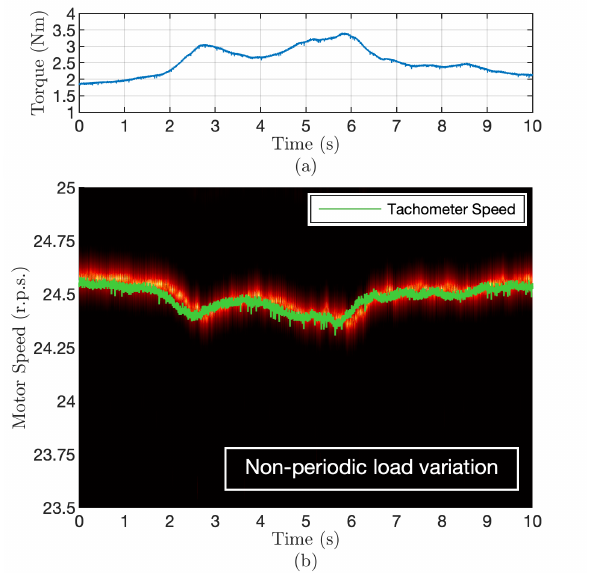}}
\caption{Experimental results: induction motor operating under a non-constant load. (a) Time-domain waveform of the load torque. (b) Speed estimation based on the sound signal recorded by the microphone.}
\label{fig_er4}
\end{figure}

\begin{figure}[htbp]
\centerline{\includegraphics[width=0.46\textwidth]{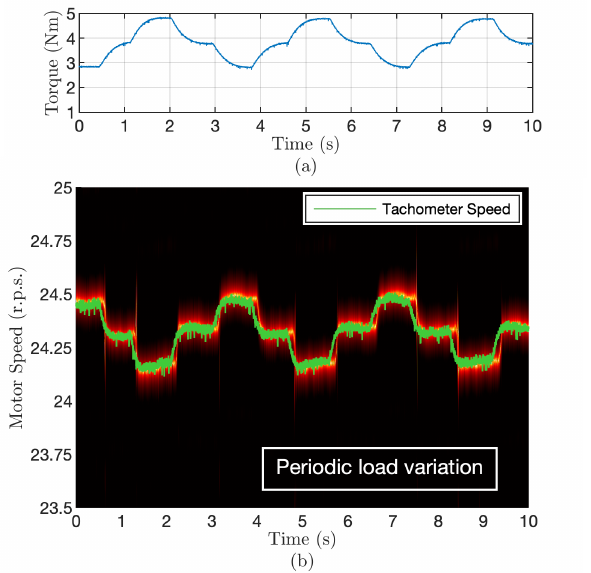}}
\caption{Experimental results: induction motor operating under a load oscillation. (a) Time-domain waveform of the oscillating load torque. (b) Speed estimation based on the sound signal recorded by the microphone.
}
\label{fig_er5}
\end{figure}

To further assess the robustness of the proposed method under dynamic operating conditions, Figs.~\ref{fig_er4} and~\ref{fig_er5} present two additional experimental trials designed to evaluate its performance under challenging conditions encountered in practical motor-driven applications. Fig.~\ref{fig_er4} illustrates a test in which the applied torque profile, shown in Fig.~\ref{fig_er4}(a), emulates realistic load conditions with both gradual and rapid variations.
The corresponding speed estimation, depicted in Fig.~\ref{fig_er4}(b), accurately captures instantaneous speed changes induced by the application of the external torque. Motor speed fluctuates within the range of 24.25 to 24.6~r.p.s., while even subtle, high-frequency variations, such as the rapid change observed at $t= 5.4$~s, are effectively detected.

Fig.~\ref{fig_er5} illustrates a second non-stationary and challenging test scenario. As shown in Fig.~\ref{fig_er5}(a), the applied load torque exhibits two step variations spanning a range of 2~Nm. The imposed torque profile is designed to emulate a low-frequency oscillatory behavior with discrete amplitude transitions, thereby enabling the evaluation of the proposed method under abrupt torque perturbations. The periodic nature of the applied changes, comprising ascending and descending steps, enables evaluating the method's response to speed variations in both directions. The resultant speed estimation, shown in Fig.~\ref{fig_er5}(b), demonstrates the method's ability to track the dynamic speed response to torque changes. The estimated speed exhibits variations between 24.12 and 24.5~r.p.s., with step changes of approximately 0.25~r.p.s. It is noteworthy that the proposed methodology accurately captures sudden decreases (e.g., at 0.6~s and 1.2~s) and abrupt increases (e.g., at 2.1~s and 3.2~s) in the motor’s instantaneous speed.

\begin{figure}[htbp]
\centerline{\includegraphics[width=0.46\textwidth]{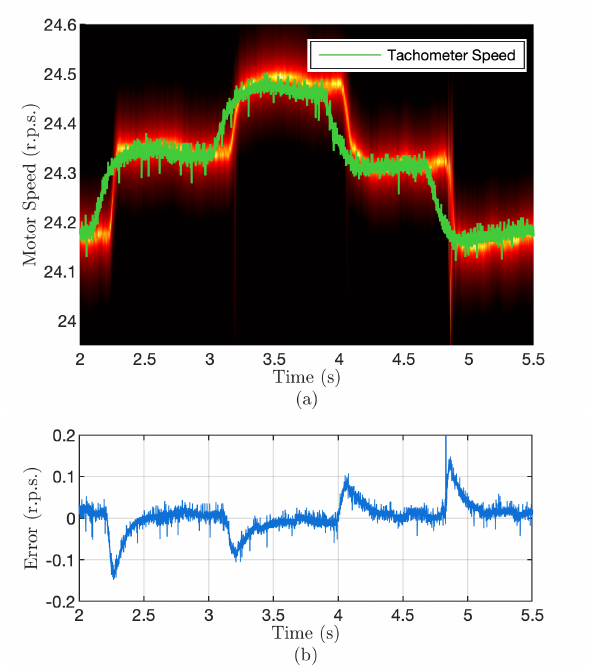}}
\caption{
Comparison between the reference and estimated rotor speed: (a) reference speed and estimated speed obtained with the proposed method. (b) corresponding estimation error, $e = n_r - \hat{n}_r$.
}
\label{fig_er6}
\end{figure}

To validate the accuracy of the proposed approach, Fig.~\ref{fig_er6}(a) compares the speed estimation results with the reference measurements obtained from the tachometer integrated into the brake. As shown, the proposed method effectively captures the dynamic variations of the motor speed, while the tachometer provides a reference signal used as ground truth. The reference signal exhibits several step-like variations, moving from 24.18~r.p.s. to 24.35~r.p.s., then to a peak near 24.49~r.p.s., before descending through 24.32~r.p.s. and settling around 24.18~r.p.s. The estimated speed closely tracks each of these transitions with minimal delay, accurately following both the rising and falling edges as well as the steady-state levels of the reference, with close agreement observed between the two traces across all operating points. 

Fig.~\ref{fig_er6}(b) shows the corresponding estimation error, defined as $e = n_r - \hat{n}_r$, over the same operating interval, which includes several step changes in the reference speed. During steady-state intervals, the error remains tightly bounded within approximately $\pm 0.02$~r.p.s. It is worth noting that the maximum peaks observed during these intervals correspond to high-frequency noise present in the tachometer signal, which further confirms the accuracy and low-noise behavior of the proposed estimator under constant-speed operation. As expected, transient peaks of up to $\pm 0.1$~r.p.s. appear at each speed-reference transition (e.g., near $t=2.3$~s, $3.2$~s, $4.1$~s, and $4.9$~s), which are primarily attributable to the intrinsic estimation delay rather than to a loss of tracking accuracy. These transients decay rapidly, with the error returning to its steady-state band within a few hundred milliseconds, demonstrating that the proposed technique preserves a fast dynamic response while maintaining high steady-state estimation precision. The frequency-domain transformation introduces an inherent time delay, since multiple samples are required to estimate the harmonic location; however, this delay is relatively small (1~ms). This delay arises from the difference between instantaneous sensor measurements and time-windowed sound signals.

To further evaluate the accuracy of the proposed acoustic-based method, the estimated rotor speed was compared with the ground-truth signal, this time omitting the delay in order to provide quantitative performance metrics. For this purpose, both signals were time-aligned and resampled to a common sampling rate to enable a point-by-point comparison. The evaluation considered periods of steady-state operation, load transitions, and transient step events. Global performance indicators, including the root-mean-square error (RMSE), mean absolute error (MAE), and mean absolute percentage error (MAPE), were computed to quantify the estimator's accuracy. The results yield an RMSE of 0.030898~r.p.s. and a MAE of 0.018881~r.p.s., indicating a high level of accuracy under realistic operating conditions. Additionally, a MAPE of 0.0776\% was obtained, confirming the reliability of the estimator across the evaluated operating range. These low error values demonstrate the capability of the proposed method to provide precise speed estimation, making it suitable for real-time monitoring and condition-monitoring applications in induction motor systems.
In general, the proposed method demonstrates a strong capability to isolate the speed-dependent harmonic component from the noise content of the acoustic signal. Moreover, the methodology effectively estimates the instantaneous speed of an electric motor under non-stationary operating conditions within typical operational ranges of industrial motors. It should be noted that the tachometer signal used as a reference is not free of error: analog and encoder-based speed transducers are affected by sensor non-idealities, quantization, and electromagnetic interference, all of which introduce measurement noise into the ground-truth speed. Therefore, a portion of the observed deviation between the acoustic estimate and the tachometer reference reflects uncertainty in the reference itself rather than error in the proposed method.

Subspace-based decomposition methods inherently involve higher computational complexity compared to conventional frequency estimators based on the FFT, particularly when high-resolution estimation is required. However, the proposed methodology incorporates an appropriate preprocessing stage within the analysis chain, which significantly reduces both computational time and memory requirements associated with the subspace matrix decomposition. As a result, the estimation delay is minimized to approximately 0.001~s, which is sufficiently fast for modeling typical electromechanical motor–load systems.
In addition, whereas conventional model-based estimation techniques require multiple electrical measurements, such as stator voltages and currents, the proposed method relies solely on a single acoustic signal. This characteristic simplifies the sensing infrastructure and facilitates practical implementation.
More importantly, the proposed approach is not sensitive to parameter variations or inaccuracies in motor parameter estimation, which commonly affect model-based methods.
A limitation of the proposed method is the requirement for prior knowledge of the number of fan blades in the motor. Nevertheless, this information is typically accessible and can be easily verified.

\section{Conclusion}
This study presents, for the first time, a subspace decomposition-based technique for estimating the rotational speed of a three-phase induction motor using acoustic measurements acquired from a low-cost microphone. The experimental results validate the effectiveness of the proposed method under both steady-state and dynamically varying load conditions. In particular, the approach enables reliable estimation of the instantaneous motor speed, accurately tracking transient variations induced by load changes as well as inherent mechanical irregularities. The adopted time--frequency subspace decomposition effectively isolates the speed-dependent harmonic component from noise and interfering spectral contributions, thereby providing meaningful insight into the motor’s dynamic behavior. Although a slight temporal delay is observed with respect to direct sensor-based measurements, this does not significantly compromise the feasibility of real-time monitoring when implemented on dedicated hardware. Overall, the proposed methodology constitutes a promising and cost-effective solution for non-invasive motor condition monitoring, demonstrating robust performance across a wide range of practical operating scenarios.

As future work, the implementation of the proposed algorithm on dedicated digital hardware platforms is envisaged to enable real-time operation. Furthermore, while the method exhibits strong performance within the typical efficient operating region of industrial motors, additional investigation is required to address more challenging regimes such as startup and shutdown transients. In this regard, future research will focus on the integration of advanced signal processing techniques to achieve accurate speed estimation during these highly non-stationary operating conditions.

%

\bibliographystyle{IEEEtran}
\bibliography{references}

\vfill

\end{document}